\documentclass{svjour2}                    
\usepackage{graphicx}
\usepackage{amsfonts, graphicx, stmaryrd, cite}
\begin{document}

\title{Ground State Energy of the One-Dimensional Discrete Random Schr\"{o}dinger Operator with Bernoulli Potential}

\author{Michael Bishop \and Jan Wehr}

\institute{M. Bishop 
	\at Department of Mathematics, University of Arizona, 617 N. Santa Rita Ave., P.O. Box 210089, Tucson, AZ 85721-0089 USA \\\email{mbishop@math.arizona.edu}\\
	J. Wehr
	\at Department of Mathematics, University of Arizona, 617 N. Santa Rita Ave., P.O. Box 210089, Tucson, AZ 85721-0089 USA \\\email{wehr@math.arizona.edu}}

\date{Received: date / Accepted: date}

\maketitle

\begin{abstract}
In this paper, we show the that the ground state energy of the one-dimensional Discrete Random Schr\"{o}dinger Operator with Bernoulli Potential is controlled asymptotically as the system size $N$ goes to infinity by the random variable $\ell_N$, the length the longest consecutive sequence of sites on the lattice with potential equal to zero.  Specifically, we will show that for almost every realization of the potential the ground state energy behaves asymptotically as
${\frac{\pi^2}{(\ell_N +1)^2}} $
in the sense that the ratio of the quantities goes to one.
\end{abstract}

\section{Introduction}

Random Schr\"odinger operators are a vast area of research in physics and mathematics.  Since the seminal work of Anderson~\cite{Anderson58} and Mott~\cite{Mott68}, there has been an uninterrupted flow of ideas and publications about their spectral properties, localization, delocalization, and dynamical properties of wave packets in a random potential etc.  The recent experimental realization of Anderson localization in cold atom systems added to the interest and motivated new research on the nonlinear random Schr\"odinger equation~\cite{Anderson1Dexp} \cite{Anderson3Dexp}.  A good general reference on the physics of quantum propagation in random media is~\cite{Lewenstein12}.  ~\cite{Kirsch07}  is an excellent survey of the mathematical theory.  A particularly important aspect of the theory is eigenvalue distribution.  We refer the reader to~\cite{Minami96} for a major result in this direction.\\

In this paper, we focus specifically on the properties of the ground state of the one-dimensional Schr\"odinger operator with a potential generated by a sequence of independent, identically distributed Bernoulli random variables, taking values $0$ and $b > 0$.  Studying the system in finite volume, we obtain detailed asymptotics of both the distribution of the smallest eigenvalue and the properties of the ground state eigenfunction as the system's size goes to infinity.  Other authors have studied different aspects of Schr\"odinger operators with Bernoulli random potentials.  In particular,~\cite{CarmonaKlein87} contains a proof of exponential localization in such systems.  We stress that from the point of view of many standard techniques, the Bernoulli case is technically difficult, because the distribution of the random potentials lacks density.  In contrast, for our direct approach, the discreteness of the potential's distribution is a simplifying feature.  It also makes it easier to compare the theoretical results to simulations in which using Bernoulli random variables is an advantage.  Roughly speaking, we prove that the ground state energy is approximately equal to the sine wave on the longest interval on which the potential equals zero and control the corrections by explicit bounds.  We emphasize that our method of estimating the ground state energy (especially from below, which is more difficult) is different from that applied e.g. in ~\cite{Simon85}.  Simon's estimate is more generally applicable, but also less accurate while our upper and lower bounds coincide up to lower order terms.  Accurate Lifschitz tail asymptotics, based on estimates similar to ones used here, are the content of a work in preparation with  V. Borovyk.  In future work with J. Stasi\'nska we will address the analog of this result for higher-dimensional systems.  ~\cite{Sznitman2011} and~\cite{Antal} are related references on diffusions processes in potentials given by Poisson clouds.  The work is motivated in part by research of the authors on the nonlinear case ~\cite{ICFO2011}.  In the remainder of the introduction, we define precisely the model and state our main result.  Its proof relies on two bounds, which are proven in the following two sections.  The proof is completed in Section 4, which also includes a limit theorem for a quantity closely related to the ground state energy.  Finally, Section 5 contains a summary and a short discussion of further research.\\

The goal of this paper is to study the ground state energy $E_0^N$ of the operator $H = -\Delta + V$ on the space $\ell^2\{1, ..., N\}$.  The Laplacian operator $-\Delta$ is considered with Dirichlet boundary conditions at the sites $0$ and $N+1$ and defined by $-\Delta f(i) = 2f(i)-f(i+1)-f(i-1)$ and $V$ is an $N$-vector with components $V(j)$.  We assume that $V(j)$ are i.i.d. random variables on a probability space $(\Omega, {\cal F}, P)$ with the Bernoulli distribution: 

$$
V(j) = \left\{ \begin{array}{l}
0\; \mathrm{\hbox{with probability}}\, p \\
b\; \mathrm{ \hbox{with probability}}\, q = 1-p
\end{array}\right.
$$
The ground state energy $E_0^N$ is defined as the minimum of $\langle \phi , H\phi \rangle$ over $\phi \in \ell^2\{1, ..., N\}$ with $\|\phi\| = 1$ (equivalently, one could take the minimum over $\phi \in \ell^2\{0, 1, ..., N, N+1\}$ with $\|\phi\| = 1$ satisfying the Dirichlet boundary conditions).  We will show that $E_0^N$ is approximately the energy of the discrete sine wave $S_N(j) = \frac{2}{\ell_N }\sin(\frac{\pi j}{\ell_N +1})$ supported on the longest island of zero potential.  This is motivated physically by the idea that finite potential barriers act essentially as infinite barriers for states with energy that converges to zero.  

The precise definitions are as follows.  The longest island of zero potential is the longest set of consecutive points $j \in \{1, \dots, N\}$ such that $V(j) = 0$.  We denote its length $\ell_N$.  There may be more than one island of length $\ell_N$, but the point is inconsequential to this paper.  The operator $H = -\Delta + V$ is self-adjoint, hence for every system size $N$ and for every realization of the random potential there exists a ground state vector $\Psi_N$ realizing the ground state energy.   Any $\omega \in \Omega$ gives rise to a realization of the potential which defines islands where the potential variables equal zero.  We denote the (consecutive) islands by $I_i$ and the length of $I_i$ by $L_i$.  The number of islands is a random variable denoted $n$ (we emphasize that while $N$ is deterministic, $n$ is random---this will be important in Section $4$).  It is natural to expect that $\ell_N \to \infty$ as $N \to \infty$ with probability $1$---this is shown in Section $4$.  We can choose $\Psi_N$ to be real and nonnegative since it is the ground state of a nonnegative operator.  (Here is the well-known argument:  the energy can be written as $\langle \Psi_N , H\Psi_N \rangle = \sum_j V(j)|\Psi_N(j)|^2 + |\Psi_N(j+1) - \Psi_N(j)|^2$.  If $\Psi_N(j+1)$ and $\Psi_N(j)$ are off by a phase factor, the energy can be made smaller by a complex phase factor correction.  This contradicts the energy-minizing property of $\Psi_N$, therefore we can assume that $\Psi_N(j)$ are all of the same complex phase factor and we will take them to be real and nonnegative.)

As a remark, our results naturally extend to the top of the spectrum: the highest eigenvalue $E^N_N$ behaves as $b - \frac{\pi^2}{(\ell_n'+1)^2}$ where the roles of $p$ and $1-p = q$ are interchanged, so $\ell_N'$ is the longest island of $b$ potential.  To prove this, we conjugate $H$ with the unitary operator $U: \phi(j) \to  (-1)^{|j|}\phi(j)$, which transforms the Laplacian spectrum to $4 + \Delta$ and leaves the potential unchanged, so the bottom of the spectrum of $b - V - U^{-1}(-\Delta) U$ is the top of the spectrum of $-\Delta + V$.  The claim now follows from our ground state result with $p$ replaced by $1-p$. ~\cite{Simon85}\\

Our main result, indicated informally above, is:\\

{\bf Theorem 1:} With probability one
$$
\lim_{N \to \infty} \frac{E_0^N}{(\frac{\pi^2}{(\ell_N +1)^2})} = 1.
$$
\section{Upper Bound on $E^N_0$}

{\bf Lemma 1}: For any realization of the potential $V$ and for any system size $N$, 
$$E_0^N \leq \frac{\pi^2}{(\ell_N +1)^2}$$.  \\
\indent	Proof:  By definition, $E_0^N = \min \{<H\phi , \phi> : \|\phi\| =1\}$.  To find an upper bound, we will construct a test function $\phi$ supported on the longest island, whose energy equals $ \sin^2(\frac{\pi}{\ell_{N}+1})$, which is of course bounded by the right-hand side of the claimed inequality.  Restriction to functions supported on the longest island can be understood as a comparison to the system with the infinite potential barrier on the island's boundaries---see the discussion in the Conclusion.  In the remainder of the proof, we will show that with this restriction the kinetic energy is minimized by  $S_N(j) = \frac{2}{L}\sin(\frac{\pi j}{L+1})$.

Consider a wave function on 
$\{0, \dots, L\}$ (treated as  $\mathbb{Z} /(L+1)\mathbb{Z}$), a cyclic group of order $L+1$, see~\cite{Stein-Shakarchi}) with Dirichlet boundary conditions: $\phi (0) = \phi (L+1) = 0$ (identifying the point $L+1$ with $0$).  Its Finite Fourier Transform is 
	$$ \widetilde{\phi}(p) =\frac{1}{\sqrt{L+1}} \Sigma_{j=0}^{L} \phi(j)e^{\frac{2\pi ipj}{L+1}} $$
The Finite Fourier Transformation diagonalizes the Discrete Laplacian.  Taking the Finite Fourier Transform of the Discrete Laplacian acting on a vector $\phi$, we obtain
	$$ \widetilde{(-\Delta\phi)}(p) = \frac{1}{\sqrt{L+1}} \Sigma_{j=0}^{L} [2\phi(j) - \phi(j-1) - \phi(j+1)]e^{\frac{2\pi ipj}{L+1}} $$
	$$= \frac{1}{\sqrt{L+1}} [2\Sigma_{j=0}^{L} \phi(j)e^{\frac{2\pi ipj}{L+1}} - \Sigma_{j=-1}^{L-1} \phi(j)e^{\frac{2\pi ip(j+1)}{L+1}} - \Sigma_{j=1}^{L+1} \phi(j)e^{\frac{2\pi ip(j-1)}{L+1}} $$

	Using the fact that $\phi (0) = \phi (L+1) = 0$ and $\phi (-1) = \phi(L)$:
	$$ = \frac{1}{\sqrt{L+1}} [2\Sigma_{j=0}^{L} \phi(j)e^{\frac{2\pi ipj}{L+1}} - \Sigma_{j=0}^{L} \phi(j)e^{\frac{2\pi ip(L+1)}{L+1}}
	- \Sigma_{j=0}^{L} \phi(j)e^{\frac{2\pi ip(j-1)}{L+1}} $$
	$$=\frac{1}{\sqrt{L+1}}[2-e^{\frac{2\pi ip}{L+1}} - e^{-\frac{2\pi ip}{L+1}}]\Sigma_{j=0}^{N-1} \phi(j)e^{\frac{2\pi ipj}{L+1}} $$
	$$=[\sin^2(\frac{p\pi}{L+1})] \widetilde{\phi}(p)$$

	This implies that the eigenvalues of $-\Delta$ are 
	$\sin^2(\frac{p\pi}{L+1})$; $p = 0, \dots L$.
	The smallest eigenvalue of the Laplacian without any boundary conditions comes from $p=0$, but the corresponding eigenfunction, which is a constant, vanishes because of the Dirichlet boundary conditions.  Therefore, the minimal energy is given by $p=1$ and $p=L$.  Taking a linear combination of the vectors $e^{\frac{i\pi}{L+1}}$ and $e^{\frac{-i\pi}{L+1}}$, corresponding to $p=1$ and $p=L$, we obtain a nonnegative vector that satisfies the Dirichlet boundary conditions:  $\phi(j) = \sin(\frac{\pi j}{L+1})$.  We calculate its norm:

	$$\|\sin(\frac{\pi j}{L+1})\|^2 = \sum_{j=1}^L|\sin^2(\frac{\pi j}{L+1})|^2  $$
	$$\sum_{j=1}^L |\sin^2(\frac{\pi j}{L+1})|^2 = \sum_{j=1}^L \frac{1}{2} - \frac{\cos(\frac{\pi j}{L+1})}{2} $$
	$$=\frac{L}{2}$$

Thus the normalized eigenvector is $S_{L+1}(j) = \sqrt{\frac{2}{L}}\sin(\frac{\pi j}{L+1})$.  (Note that in the above we considered vectors defined on the cyclic group $\mathbb{Z} /(L+1)\mathbb{Z}$ rather than on $\{0, ... ,  L+1\}$.   However, since the discrete sine wave is zero at the boundary points, the inner product will not include the boundary terms of the Laplacian acting on the vector.)  \\

We define the test function $\psi(j)$ on $\{0, 1, ... , N+1\}$ as follows.  On the longest island of zero potential, let $\psi(j) = S_{\ell_{N+1}}(j)$ and let $\psi(j)=0$ for all other $j$.  The energy of this test function is
	$$ \langle H \psi, \psi \rangle $$
	$$= \langle -\Delta S_{\ell_{N}+1},S_{\ell_{N}+1}\rangle $$
	$$ = \sin^2(\frac{\pi}{\ell_{N}+1}) $$
	$$ \leq \frac{\pi^2}{(\ell_{N}+1)^2}$$
The ground state energy is defined as the minimizer of $<H\phi , \phi>$ for all norm one vectors.  Therefore, the ground state energy must be less than the energy of the test function.
	
\hfill $\Box$

\section{Lower Bound on $E^N_0$}

For each realization of potential and for every $N$, we have a specific nonnegative ground state $\Psi_N$.  In this section, we prove a lower bound on the energy of $\Psi_N$, which will be our main technical tool.  This bound approaches the upper bound of the previous section when $\ell_N \to \infty$ as $N\to \infty$.  We will show in section $5$ that this is true with probability one.  We define $\delta^i_L$ and $\delta^i_R$ as the values of $\Psi_N$ on the sites with potential equal to $b$, adjacent to the left and the right of the island $I_i$ and define $m_i$ as the norm of $\Psi_N$ on $I_i$, $m_i^2 = \|\Psi_N|_{I_i}\|^2$.   We will use the definition of $\Psi_N$ as the energy minimizer to estimate its energy from below.  As the critical tool, we first estimate its norm on sites of $b$ potential, using the the upper bound on the ground state energy proven in Section 3.  

\vspace{.5cm}

{\bf Corollary 1:} Let $\Psi_N$ be the ground state vector for a given realization of potential.  Let $B = \{ j : V(j) = b\} $.  Then $\| \Psi_N|_B \|^2 \leq \frac{\pi^2}{b(\ell_N+1)^2}$ by Lemma 1, where $\Psi_N|_B$ is the restriction of $\Psi_N$ to the set $B$. \\
\indent Proof: The contribution of energy of $\Psi_N$ on $B = \{j: V(j)=b\}$ is bounded by the quantity in Lemma 1.
	$$\frac{\pi^2}{(\ell_N+1)^2}  $$
	$$ \geq  \langle H\Psi_N,\Psi_N \rangle  $$
	$$\geq  \langle V\Psi_N|_B, \Psi_N|_B\rangle $$
	$$= b\|\Psi_N|_B\|^2$$
Therefore, $\|\Psi_N|_B\|^2 \leq \frac{\pi^2}{b(\ell_N+1)^2}$.  
\hfill $\Box$

It follows from the corollary that $ \sum_{i=1}^n (|\delta_L^i|^2 + |\delta_R^i|^2) \leq \frac{2\pi^2}{b(\ell_N+1)^2}$  where the extra factor of 2 is introduced to cover the case in which the outer boundaries are shared by islands.  The norm of $\Psi_N$ is thus bounded below on sites of zero potential by $\|\Psi_N|_{B^c}\|^2 \geq 1 - \frac{\pi^2}{b(\ell_N+1)^2}$.  

Next, we fix $\gamma \in (0,1)$.  While in what follows the specific value of the parameter $\gamma$ does not play a crucial role, it will enter the convergence rate in our main result and its choice may become important in further work based on similar ideas.\\

{\bf Definition:}  A \textbf{heavy island of $\Psi_N$} is an island $I_i$ such that $\|\Psi_N|_{I_i}\|^2 \geq \max(\delta_L^i,\delta_R^i)^2(L_i+1)(\ell_N+1)^{1-\gamma}$.  The island which is not heavy is (naturally) called \textbf{light}.\\

For the lower bound on the energy, we will estimate the contributions of the heavy islands  of $\Psi_N$ and neglect the contributions of the light islands.  First, we show that norm of $\Psi_N$ is concentrated on the heavy islands.\\ 

{\bf Lemma 2:} Let $M$ be the index set of all heavy islands $I_i$.  Then $$\Sigma_{i \in M} \|\Psi_N|_{I_i}\|^2 \geq 1 -  \frac{3\pi^2}{b\ell_N^\gamma}$$  \\ 

\indent Proof: 
First, we split $\Psi_N$ into two orthogonal vectors, where $B = \{j: V(j)=b\}$.
$$1 = \|\Psi_N\|^2 = \|\Psi_N|_B\|^2 + \|\Psi_N|_{B^c}\|^2$$
Next, we split $\Psi_N|_{B^c}$ into heavy islands and light islands.
$$ \|\Psi_N|_{B^c}\|^2 = \Sigma_{i \in M} \|\Psi_N|_{I_i}\|^2 + \Sigma_{i \in M^c} \|\Psi_N|_{I_i}\|^2$$
By the definition of light islands, their contribution to the norm of $\Psi_N$
is bounded above.
	$$ \Sigma_{i \in M^c} \|\Psi_N|_{I_i}\|^2 \leq  \Sigma_{i \in M^c} \max(\delta_L^i,\delta_R^i)^2  (L_i+1)(\ell_N+1)^{1-\gamma}$$
	$$\leq  \Sigma_{i \in M^c} \max(\delta_L^i,\delta_R^i)^2 (\ell_N+1)^{2-\gamma} $$
	$$\leq \frac{2\pi^2}{(\ell_N +1)^\gamma}$$
where $\Sigma_{i \in M^c} \max(\delta_L^i,\delta_R^i)^2 \leq 2\|\Psi_N|_B\|^2 \leq  \frac{2\pi^2}{b(\ell_N+1)^2}$.   Thus
	$$1 = \|\Psi_N|_B\|^2 + \Sigma_{i \in M} \|\Psi_N|_{I_i}\|^2 + \Sigma_{i \in M^c} \|\Psi_N|_{I_i}\|^2 $$
	$$\leq  \frac{\pi^2}{b(\ell_N +1)^2} + \frac{2\pi^2}{(\ell_N +1)^\gamma} + \Sigma_{i \in M} \|\Psi_N|_{I_i}\|^2 $$
	$$\leq \frac{3\pi^2}{b(\ell_N + 1)^\gamma}+ \Sigma_{i \in M} \|\Psi_N|_{I_i}\|^2$$
Subtracting appropriate terms, the bound is shown.
\hfill $\Box$\\

The lemma guarantees we have at least one heavy island when $\frac{3\pi^2}{b(\ell_N + 1)^\gamma} < 1$.  The following theorem is the main technical result that bounds the energy below by estimating the contribution of energy of $\Psi_N$ on the heavy islands.\\

{\bf Lemma 3:}  If $\ell_N > \frac{3\pi^2}{b}^{\frac{1}{\gamma}} - 1$ , then the lower bound on the energy contributed by heavy islands is
$$\left(1 - \frac{3\pi^2}{b\ell_N^\gamma} \right) \left( 1 - \frac{1}{\ell_N^{(1-\gamma)/2}}\right)^2 \left( \frac{\pi^2}{(\ell_N+1)^2}  + O(\frac{\pi^4}{(\ell_N+1)^4})\right)$$
and thus $E_0^N$ is also bounded below by this quantity.\\

\indent Proof:  We will first find an explicit form of the state $\Psi_N$ minimizing the energy on a heavy island $I_i$ (which exists by the condition of the theorem) using the Lagrange multiplier method with fixed nonnegative values of $\delta_L^i$ and $\delta_R^i$ as well as $m_i$---the norm of the restriction of the state to the $i$-th island.  Afterwards, we will obtain information about the values of $m_i$, $\delta_L^i$ and $\delta_R^i$ in the actual ground state.

To find the energy-minimizing vector under these conditions, consider, for a fixed realization of the $V$, the function $E = \langle f,Hf\rangle = \Sigma_{j=1}^{L} (-\Delta f(j)) f(j)$ subject to the constraint $\Sigma_{j=1}^{L_i} f(j)^2 = m_i^2$ and  $f(0) = \delta_L^i$ and $f(L+1) = \delta_R^i$, the boundary conditions set by $\Psi_N$.  Because the components of the minimizing vector are nonnegative, we only need to consider the vectors with real compoments.

Introducing a Lagrange multiplier $\lambda$, the minimizing vector satisfies
  $$\nabla \langle f,Hf\rangle  = \lambda \nabla \Sigma_{j=1}^{L_i} f(j)^2$$ 
which gives the set of equations.
$$ \frac{\partial E}{\partial f(1)} = 4f(1) - 2\delta_L^i - 2f(2)  = 2\lambda f(1)$$
$$ \frac{\partial E}{\partial f(j)} = 4f(j) - 2f(j-1) - 2f(j+1) = 2\lambda f(j)  $$
$$ \frac{\partial E}{\partial f(L)} = 4f(L) - 2\delta_R^i - 2f(L-1) = 2\lambda f(L) $$
With some rearranging, we obtain the eigenvalue equation
$$ -\Delta f(j) = \lambda f(j)$$
The critical points  are eigenvectors of $-\Delta$ with boundary conditions $f(0) = \delta_L^i$ and $f(L+1) = \delta_R^i$.  The set of eigenvectors contains $f_0(j) = \frac{\delta_R-\delta_L}{L_i+1} j + \delta_L$ and $f_k(j) = c_k \sin(s_k \frac{k\pi}{L_i+1}j + t_k)$, where $k = 1, ..., L$ is the frequency index, $c_k$ is a normalizing constant, and $s_k$ and $t_k$ make the sine wave satisfy the $f(0) = \delta_L^i$ and $f(L+1) = \delta_R^i$ boundary conditions.  Since $I_i$ is a heavy island, $\|\Psi_N|_{I_i}\|^2 \geq \max(\delta_L^i, \delta_R^i)^2L_i(\ell_N+1)^{1-\gamma}$.  The square of the norm of the linear function $f_0$ satisfies $\|f_0\|^2 \leq \max(\delta_L^i,\delta_R^i)^2 L_i$ and is strictly smaller than $\max(\delta_L^i,\delta_R^i)^2L_i(\ell_N+1)^{1-\gamma}$ , so the minimizing vector must be one of the sine waves.  Since $\Psi_N$ is strictly positive, $s_k k$ must be sufficiently close to $1$ (we will prove a precise estimate later) and it is notationally convenient to put $k=1$, so that $f_1$ is the minimizing vector.  Evaluating $\langle Hf_1, f_1\rangle  =  \langle -\Delta c_1 \sin(s_1 \frac{\pi}{L_i+1}j + t_1),c_1 \sin(s_1 \frac{1\pi}{L_i+1}j + t_1)\rangle = m_i^2\frac{s_1^2\pi^2 }{(L_i+1)^2}  + O(\frac{s_1^4\pi^4}{(L_i+1)^4})$ gives us the energy contribution of $\Psi_N|_{I_i}$.\\

To estimate the energy of the ground state, we must understand how the coefficients $s_1$, $t_1$ and $c_1$ depend on the boundary conditions $\delta_L^i$ and $\delta_R^i$ and the mass $m_i$.  In particular, we will find a precise estimate on $s_1$.  The frequency index $1$ is dropped from $f_1$, $s_1$, $t_1$, and $c_1$ for simplicity.  The boundary conditions are:

$$ c \sin(t) = \delta_L^i$$
$$ c \sin(s\pi+t) = \delta_R^i$$

Note that both $\delta_L^i$ and $\delta_R^i$ are positive and smaller than $\frac{\pi^2}{b(\ell_N + 1)^2}$.  Since $I_i$ is a heavy island, $f$ cannot be monotone on all of it because in this case its mass there would be bounded above by $\max(\delta_L^i, \delta_R^i)^2L_i$, which would contradict the definition of a heavy island.  Since the ground state is nonnegative, it follows that $s$ is bounded above by $1$ .  To solve for $s$, we use the left boundary condition to solve for $t$ where $t \in [0, \pi/2)$.
$$t = \arcsin (\frac{\delta_L^i}{c})$$

We deal with the right boundary condition similarly. As argued above, $f$ is not monotone on $I_i$, so solving the right boundary condition for $s\pi + t$ requires taking the inverse of $ y= \sin(x)$ on $[\pi/2, 3\pi/2]$ which is $x = -\arcsin (y) + \pi$
$$s\pi + t = -\arcsin(\frac{\delta_R^i}{c}) + \pi $$
Subtracting the two equations we get the following equation for $s$.
\begin{equation} s = 1 - \frac{1}{\pi}(\arcsin(\frac{\delta_L^i}{c}) + \arcsin(\frac{\delta_R^i}{c})) \end{equation}
To bound $s$ from below, $\frac{\delta_R^i}{c}$ and $\frac{\delta_L^i}{c}$ need to be bounded above.  To do this, calculate that the square norm of the vector restricted to the island before normalization.   
$$\| \sin(s\frac{\pi}{L_i+1}x + t)|_{I_i} \|^2 = \Sigma_{j=1}^{L_i} \sin^2(s\frac{\pi}{L_i+1}x + t)$$
\begin{equation}= \frac{L_i}{2} - \frac{L_i+1}{2}\Sigma_{j=1}^{L_i} \frac{\cos(2(s\frac{\pi}{L_i+1}j + t))}{L_i+1}\end{equation} 
The second term is a Riemann sum of the function $\cos(2(s\pi x + t))$ multiplied by $\frac{L_i + 1}{2}$.  The associated integral is simpler to calculate.  The difference between the sum and integral is bounded by
\begin{equation}\left| \Sigma_{j=1}^{L_i+1} \frac{\cos(2(s\frac{\pi}{L_i+1}j + t))}{L_i+1} - \int_0^1 \cos(2(s\pi x + t)) dx \right| + \frac{1 - 2(\delta_R^i)^2}{L_i+1}\end{equation}
where the second term is the missing term from the Riemann sum (the right endpoint).   We can estimate the difference between the Riemann sum and the integral by noting that the derivative of $\cos(2(s\pi x +t))$ is bounded by $2s\pi$.  Therefore the expression (3) is bounded above the following error term.
$$ \Sigma_{j=1}^{L_i} \frac{2s\pi + 1 - 2(\delta_R^i)^2}{(L_i+1)^2} = \frac{2s\pi + 1- 2(\delta_R^i)^2}{L_i+1}$$

The Riemann sum is bounded below by the integral minus the error term.  Continuing with the calculation of the square norm, expression (2) is bounded above by:
\begin{eqnarray}
	&& \frac{L_i}{2} - \frac{L_i+1}{2}\int_0^1 \cos(2(s\pi x + t)) dx + s\pi + 1/2 - (\delta_R^i)^2 \nonumber\\
	&=& \frac{L_i + 1}{2} - \frac{L_i+1}{4s\pi} (\sin(2(s\pi + t)) - \sin(2t)) + s\pi - (\delta_R^i)^2 \nonumber\\
	&=& \frac{L_i+1}{2} - \frac{L_i+1}{4s\pi} (2\sin(s\pi + t)\cos(s\pi + t) - 2\sin(t)\cos(t)) \nonumber\\
	&+& s\pi  - (\delta_R^i)^2 
\end{eqnarray}
Now, we can substitute using the boundary equations.  Since $s\pi + t \in [\frac{\pi}{2},\pi ]$, $c\cos(s\pi + t) = -\sqrt{c^2 - (\delta^i_R)^2}$ and since $t \in [0,\frac{\pi}{2}]$, $c\cos(t) = \sqrt{c^2 - (\delta_L^i)^2}$.  Therefore, the expression (4) is equal to 
$$\frac{L_i + 1}{2}\left[ 1  + \frac{1}{ s\pi}\left(\frac{\delta^i_R}{c}\sqrt{1 - \frac{(\delta^i_R)^2}{c^2}} + \frac{\delta_L^i}{c}\sqrt{1 - \frac{(\delta^i_L)^2}{c^2}}\right)\right] + s\pi - (\delta_R^i)^2$$
This is a bound for $\|c\sin(\frac{s\pi x}{L_i + 1} + t)\|^2$.  By definition, $\| f\|^2 = \| c \sin(s\frac{\pi}{L_i+1}\cdot +t)\|^2$, so
\begin{eqnarray}
	&&  \|f\|^2 \leq \frac{L_i + 1}{2}\left[c^2 + \frac{1}{ s\pi} \left( \delta^i_R\sqrt{c^2 - (\delta^i_R)^2} + \delta^i_L\sqrt{c^2 - (\delta^i_L)^2}\right)\right] \nonumber\\
	&& \ \ \ \ \ \ \ \ \ \ + c^2(s\pi - (\delta_R^i)^2) 
\end{eqnarray}

The last equation can be solved for the first term in the square brackets $c^2$.  We then divide by the larger of the values $\delta_L^i$ and $\delta_R^i$.  Without loss of generality, assume $\delta_L^i \leq \delta_R^i$.
$$ \frac{c^2}{(\delta^i_R)^2} \geq  -\frac{1}{s\pi} \left( \sqrt{\frac{c^2}{(\delta^i_R)^2} - 1} +\frac{ \delta^i_L}{\delta^i_R}\sqrt{\frac{c^2}{(\delta^i_R)^2} - \frac{(\delta^i_L)^2}{(\delta^i_R)^2}}\right) + \frac{2\|f\|^2}{(L_i+1)(\delta^i_R)^2} - \frac{2c^2(s\pi - (\delta_R^i)^2)}{(L_i+1)(\delta_R^i)^2} $$
Move the last term to the left side to get the inequality
$$ \frac{c^2}{(\delta^i_R)^2} \left( 1 + \frac{2(s\pi - (\delta_R^i)^2)}{L_i + 1}\right) \geq   $$
\begin{equation}- \frac{1}{s\pi}\left( \sqrt{\frac{c^2}{(\delta^i_R)^2} - 1} +\frac{ \delta^i_L}{\delta^i_R}\sqrt{\frac{c^2}{(\delta^i_R)^2} - \frac{(\delta^i_L)^2}{(\delta^i_R)^2}}\right) +  \frac{2\|f\|^2}{(L_i + 1)(\delta^i_R)^2} \end{equation}

Since the island $I_i$ is  heavy, $\|f\|^2 = \|\Psi_N|_{I_i}\|^2 \geq \max(\delta_L^i,\delta_R^i)^2(L_i+1)(\ell_N+1)^{1-\gamma}$, and 

$$ \frac{c^2}{(\delta^i_R)^2} \left( 1 + \frac{(s\pi - (\delta_R^i)^2)}{L_i + 1}\right) \geq  $$
\begin{equation} - \frac{1}{s\pi} \left( \sqrt{\frac{c^2}{(\delta^i_R)^2} - 1} +\frac{ \delta^i_L}{\delta^i_R}\sqrt{\frac{c^2}{(\delta^i_R)^2} - \frac{(\delta^i_L)^2}{(\delta^i_R)^2}}\right) +  2(\ell_N+1)^{1-\gamma} \end{equation}

As stated before, $s$ is bounded above by $1$ and below by $0$ and the $\delta$-terms are bounded above by $\frac{\pi^2}{b(\ell_N+1)^2}$, so $\frac{(s\pi - (\delta_R^i)^2)}{L_i} \geq -\frac{\pi^2}{b(L_1+1)(\ell_N + 1)^2} > -1$ which means this term does not effect the limiting behavior of $\frac{c^2}{(\delta^i_R)^2}$ as $\ell_N \to \infty$.  Since this is a heavy island, $s$ is strictly greater than zero because $\Psi_N$ achieves a unique maximum on $I_i$ that is not an endpoint.  The limiting behavior of the negative term as $\ell_N \to \infty$ depends only on the behavior of the square root terms.  

We want to show  the term $2(\ell_N+1)^{1-\gamma}$ dominates as $\ell_N \to \infty$.  The order of 
$$\frac{1}{s\pi}\left( \sqrt{\frac{c^2}{(\delta^i_R)^2} - 1} +\frac{ \delta^i_L}{\delta^i_R}\sqrt{\frac{c^2}{(\delta^i_R)^2} - \frac{(\delta^i_L)^2}{(\delta^i_R)^2}}\right)$$ is at most the order of $\frac{c}{\delta^i_R}$.  It follows that the order of $\frac{c^2}{(\delta^i_R)^2}$  does not depend on the square root terms, so the term $2(\ell_N+1)^{1-\gamma}$ must dominate.  Therefore, 
$$\frac{c^2}{(\delta^i_R)^2} \geq  2\ell_N^{1-\gamma} - O(\ell_N^{(1-\gamma)/2})$$
 and 
$$\frac{(\delta^i_R)^2}{c^2} \leq   \frac{1}{2\ell_N^{1-\gamma} - O(\ell_N^{(1-\gamma)/2})}$$.  Because $\ell_N > 1$, $2\ell_N^{1-\gamma} - O(\ell_N^{(1-\gamma)/2}) > \ell_N^{1-\gamma}$ and we assumed $\delta^i_L \leq \delta^i_R$, we have
$$\frac{(\delta^i_L)^2}{c^2} \leq \frac{(\delta^i_R)^2}{c^2} \leq   \frac{1}{\ell_N^{1-\gamma}}$$
Since $\arcsin(x) \leq \frac{\pi x}{2}$, we can apply the bound to equation (1)
$$s = 1 - \frac{1}{\pi}(\arcsin(\frac{\delta^i_L}{c}) + \arcsin(\frac{\delta^i_R}{c}))$$
$$ \geq 1 -  \frac{\frac{\delta^i_L}{c} +\frac{\delta^i_R}{c}}{2}$$
$$ \geq 1 - \frac{1}{\sqrt{\ell_N^{1-\gamma}}} $$

For all heavy islands, $s \geq 1 - \frac{1}{\ell_N^{(1-\gamma)/2}} $.  Then the energy contribution of each of the heavy islands is
	$$m_i^2 \left( \frac{s_i^2 \pi^2}{(L_i+1)^2}  + O(\frac{\pi^4}{(L_i+1)^4})\right)$$
which is bounded below by  
$$m_i^2 \left( \frac{\left( 1 - \frac{1}{\ell_N^{(1-\gamma)/2}}\right)^2\pi^2}{(\ell_N+1)^2}  + O(\frac{\pi^4}{(\ell_N+1)^4})\right)$$
where we used $L_i \leq \ell_N$ and dropped $s_i$ from the fourth order term since $s \leq 1$.  The total mass of heavy islands is bounded below by $\left(1 - \frac{3\pi^2}{b(\ell_N+1)^\gamma} \right)$ by Lemma $2$.  Thus the energy contribution of heavy islands is bounded below by 

$$\left(1 - \frac{3\pi^2}{b\ell_N^\gamma} \right) \left( 1 - \frac{1}{\ell_N^{(1-\gamma)/2}}\right)^2 \left( \frac{\pi^2}{(\ell_N+1)^2}  + O(\frac{\pi^4}{(\ell_N+1)^4})\right)$$
Since the ground state energy is bounded below by the energy contribution of heavy islands, we have the desired lower bound.

\hfill $\Box$

Using Lemma 1 and 3, the ground state energy is controlled as 
$$\left(1 - \frac{3\pi^2}{b\ell_N^\gamma} \right) \left( 1 - \frac{1}{\ell_N^{(1-\gamma)/2}}\right)^2 \left( \frac{\pi^2}{(\ell_N+1)^2}  + O(\frac{\pi^4}{(\ell_N+1)^4})\right) \leq E_0^N \leq \frac{\pi^2}{(\ell_N +1)^2}$$
Therefore, as $\ell_N \to \infty$, $E_0^N \approx \frac{\pi^2}{(\ell_N +1)^2}$.  \\

\section{Limiting Distribution of $\ell_N$}

	This section is a brief discussion of the limiting distribution of $\ell_N$ with most details left to the reader.  It is natural to conjecture that the distribution of $\ell_N$ determines the distribution of $E_0^N$, but the bounds in the previous section are not strong enough to give that result.  The bounds do, however, show the asymptotic relation
$$P\left[\lim_{N \to \infty} \frac{E_0^N}{\frac{\pi^2}{(\ell_N +1)^2}} = 1\right] = 1 $$
claimed in Theorem 1.  We finish here the proof of this statement and, for completeness, we derive a limit theorem for $\ell_N$.  

The first step is to find the distribution of $M_j$, the maximum length of $j$ independent islands.  This uses the theorem from ~\cite{Leadbetter} that states
	$$j(1-P[X_1 \leq u_j(\tau)]) \rightarrow \tau \Rightarrow P[M_j \leq u_j(\tau)] \rightarrow e^{-\tau}$$
where $X_1$ is the individual random variable and $M_j$ is the maximum of $j$ i.i.d. random variables with the same distribution as $X_1$.  Note that the distribution of an individual island is geometric:  $P[L = k] = p^{k-1} q$, for $k=1, 2, ...$  ($k$ starts from one because an island has at least one site of zero potential).  Because $P[L \leq x] = P[L \leq \lfloor x\rfloor]$, where $ \lfloor x\rfloor$ is the largest integer smaller than $x$, for the $u_j(\tau)$to have a limit we must choose a subsequence $j_k$ such that the fractional part of $- \log_pj_k$, denoted $\{  -\log_pj_k \}$, converges monotonically from above or from below to $\theta \in [0,1]$.  When $\{ -\log_pj_k \} \downarrow \theta \in [0,1)$ we get the following limiting distribution:

$$P[ M_{j_k} \leq u_{j_k}(\tau)] \to \exp[- p^{\tau - \{ \tau + \theta \}}] + O(1/{j_k})$$
This shows the possible subsequential limits parameterized by $\theta$.

The second step is to study the number of islands $n$ in a system of size $N$.  In fact, $\frac{n}{N}$ converges to $\mu = \frac{1}{2 + \frac{1}{pq}}$ in probability.  To show this, we estimate the probabilities of the events 
$$P\left[\{ \omega : n < \lfloor (\frac{1}{\mu} - \delta) N \rfloor \}\right]$$
 and 
$$P\left[\{ \omega : n > \lfloor (\frac{1}{\mu} + \delta) N \rfloor \}\right]$$ 
by 
$$P\left[\{ \omega : \sum_{j=1}^{ \lfloor (\frac{1}{\mu} - \delta) N \rfloor +1} m_j + m' > N\}\right]$$ 
and 
$$P\left[\{ \omega : \sum_{j=1}^{\lfloor (\frac{1}{\mu} + \delta) N \rfloor} m_j  + m < N \}\right]$$ 
respectively, where $m_j$ is the sum of the lengths of one island of zero potential and one island of $b$ potential and $m$ is a correction random variable for an extra or missing island.  We apply the exponential Chebyshev inequality to the latter probabilities and use moment generating functions to derive an exponential bound.  Convergence with probability one can also be shown, but it is unnecessary for finding the limiting distribution.

The final step is to estimate $\ell_{N_k}$ above and below by $M_{\mu N_k + \delta N_k}$ and $M_{\mu N_k - \delta N_k}$ respectively in any realizations such that $|\frac{n}{N_k} - \mu | < \delta$.  The probability that $n$ does not satisfy this condition is bounded by an arbitrarily small $\epsilon$ provided $N_k$ is large enough, due to the convergence in probability.  The $M$-variables then converge (over a subsequence) to one of the limiting distributions found above, as $N_k \to \infty$.  Note that $\log_p(\mu N_k + \delta N_k) \approx \log_p(\mu N_k) + \delta'$, where $\delta'$ depends directly on $\delta$ so the convergence of the irrational rotation by the $\{ \log_p(\mu N_k)\}$ is preserved.  Taking $\delta \to 0$ we find that the limiting distribution at all points of continuity (all $\tau$ such that $\{\tau\} \neq 1 - \theta$) is given by:

 $$ P\left[\ell_{N_k}  - \frac{\log\left(\lfloor \frac{1}{\mu}N_k\rfloor\right)}{|\log p |}\leq \tau \right] \to \exp\left[-  p^{\tau - \{\tau + \theta \} }\right] $$
This is a limit theorem satisfied by (subsequences of) $\ell_N$.
As a corollary, $$P\left[ \lim_{N\to\infty} \ell_N = \infty\right] = 1$$
which combined with our bounds on $E_0^N$ gives

$$P\left[\lim_{N \to \infty} \frac{E_0^N}{\frac{\pi^2}{(\ell_N +1)^2}} = 1\right] = 1 $$
thus finishing the proof of our main result---Theorem 1.   \\

\bigskip
{\bf Remark:}  Attempting to derive a limiting distribution for $E_0^{N_k}$, a change of variables to find the distribution of $\frac{\pi^2}{(\ell_{N_k}+1)^2}$ would result in:

$$P\left[\frac{\log^3 \lfloor \mu {N_k}\rfloor}{2\pi^2 \log^2p}\left(\frac{\pi^2}{\ell_{N_k}^2} - \frac{\pi^2 \log^2p}{\log^2 \lfloor \mu {N_k}\rfloor}\right) \geq \tau \right] $$
The difference between $\frac{\pi^2}{\ell_{N_k}^2}$ and $\frac{\pi^2 \log^2p}{\log^2 \lfloor \mu {N_k}\rfloor}$ is of the order $\log^3 \lfloor \mu {N_k} \rfloor$.  The correction term on the lower bound is at best of the order $\log ^{-2 - \frac{1}{3}}(\lfloor \mu N_k\rfloor)$ which diverges when multiplied by the $\log^3\lfloor\mu N)k\rfloor$ factor.  If we try the change of variables with the lower bound, we will run into the same problem.  This shows that our estimates are not sufficient to establish a limit theorem for  the ground state energy.

\section{Conclusion}
For the one-dimensional discrete random Schr\"{o}dinger operator with Bernoulli Potential, our key result is that $E_0^N$ is controlled  in the limit as $N \to \infty$ by 
$$ \left(1 - \frac{3\pi^2}{b\ell_N^\gamma} \right) \left( 1 - \frac{1}{\ell_N^{(1-\gamma)/2}}\right)^2 \left( \frac{\pi^2}{(\ell_N+1)^2}  + O(\frac{\pi^4}{(\ell_N+1)^4})\right) \leq E_0^N $$
\begin{equation} E_0^N \leq \frac{\pi^2}{(\ell_N +1)^2}
\end{equation}
Since $P\left[ \lim_{N \to \infty} \ell_N = \infty \right] = 1$, it follows that

$$P\left[\lim_{N \to \infty} \frac{E_0^N}{\left(\frac{\pi^2}{(\ell_N +1)^2}\right)} = 1\right] = 1 $$
The key technical part of the proof is the derivation of the lower bound which uses classification of islands into heavy and light. 

As pointed out in the proof of Lemma 1, the argument used there essentially compares the given system to a system with infinite potential barriers rather than barriers of height $b$.  That we are able to prove an asymptotically equivalent lower bound means physically that, as far as the ground state energy is concerned, the two systems are not very different and consequently, its leading order behavior does not depend on $b$ as made explicit by (8).

It can be seen from the proof, that randomness of the potential is used only to show that $\ell_N \to \infty$.  The argument for both the upper and the lower bound works for individual potential realizations which satisfy this condition.  The same result thus holds for any deterministic potential with this property.  One could even consider a sequence of volume-dependent potentials (taking values $0$ and $b$).  For example, the theorem still holds for a sequence of periodic potentials whose periods diverge with $N$. We owe this remark to an anonymous referee.

Physical intution, as well as numerical results, lead us to believe that the ground state is approximately the sine wave on the longest island of zero potential.  If the ground state does localize to the largest island (or one of them), then the ground state energy calculation simplifies greatly since, in the notation of section 4, $\|f\| \approx 1 - o(1)$ can be substituted and it can be shown that $s \approx 1 - \frac{1}{\sqrt{\ell_N}}$.  We plan to address this issue in a forthcoming paper.  

We conjecture that, similar to the ground state, the low energy excited states will be approximated by half-sine waves and higher harmonics on the long islands.  This may be lead to a strong form of a Lifschitz tail estimate in the one-dimensional  discrete random Schr\"{o}dinger operators with Bernoulli potential.  Related ideas are used in work on many particle systems~\cite{ICFO2011}.

\begin{acknowledgements}

We would like to thank W. Faris for the suggestion to look at potentials with Bernoulli distributions which turned out to be a great starting point.  We would like to thank R. Sims for useful discussions, in particular of the broader context of random Schr\"dinger operator theory.  We would like to thank M. Lewenstein, A. Sanpera, P. Massignan, and J. Stasi\'nska for collaborating with us on related projects as well as providing supporting numerical results.  The first idea of the present paper grew out of discussions with J. Xin.  Both authors were supported in part by NSF grant DMS-1009508 and M. Bishop was in addition funded by NSF VIGRE grant DMS-0602173 at the University of Arizona and by the NSF under Grant No DGE-0841234.\\

\end{acknowledgements}

\bibliography{bib}
\bibliographystyle{plain}

\end{document}